\documentclass[12pt,manuscript]{aastex}
\begin{document}
\newcommand{\vdag}{(v)^\dagger}
\newcommand{\degrees} {$^\circ$}
\newcommand{\kms}{km\ s$^{-1}$}
\newcommand{\cc}{cm$^{-3}$}
\newcommand{\cotwoone} {$^{12}$CO (2-1)}
\newcommand{\co} {$^{12}$CO}
\newcommand{\thco} {$^{13}$CO}
\newcommand{\myemail}{agoodman@cfa.harvard.edu}


\shorttitle{Speeding Young Star?}
\shortauthors{Goodman \& Arce}

\title{PV Ceph: Young Star Caught Speeding?}


\author{Alyssa A. Goodman}
\affil{Harvard-Smithsonian Center for Astrophysics, Cambridge, MA 02138}
\email{agoodman@cfa.harvard.edu}

\author{H\'ector G. Arce}
\affil{California Insitute of Technology, Pasadena, CA}
\email{harce@astro.caltech.edu}


\begin{abstract}
Three independent lines of evidence imply that the young star PV Ceph is
moving at roughly 20 \kms\ through the interstellar medium.  The first, and
strongest, suggestion of motion comes from the geometry of the HH knots in
the ``giant" Herbig-Haro (HH) flow associated with PV Ceph.  Bisectors of lines
drawn between pairs of knots at nearly equal distances from PV Ceph imply an
E-W motion of the source, and a plasmon model fit to the knot positions
gives a good fit of 22 \kms\ motion for the star.  The second bit of damning
evidence comes from a redshifted ``trail" of molecular gas, pointing in the
{\it same E-W direction} implied by the HH knot geometry.  The third exhibit
we offer in accusing PV Ceph of speeding involves the tilt apparent in the
high-velocity molecular jet now emanating from the star.  This tilt is best
explained if the true, current, jet direction is N-S, as it is in HST WFPC2
images, and the star is moving--again at roughly 20 \kms.

Tracing the motion of PV Ceph backward in time, to the nearest cluster from
which it might have been ejected, we find that it is very likely to have
been thrown out of the massive star-forming cluster NGC 7023 ---more than 10 pc
away.  PV Ceph and NGC 7023 are at similar distances, and the backward-trace
of PV Ceph's motion is astonishingly well-aligned with a dark, previously
unexplained, rift in NGC 7023.  We propose that PV Ceph was ejected, at a
speed large enough to escape NGC 7023, at least 100,000 years ago, but that
it did not enter the molecular cloud in which it now finds itself until more
like 35,000 years ago.  Our calculations show that currently-observable
molecular outflow associated with PV Ceph is about 10,000 years old, so that
the flow has had plenty of time to form while in its current molecular
cloud.  But, the question of what PV Ceph was doing, and what gas/disk it
took along with it in the time it was traveling through the low-density
region between NGC 7023 and its current home is open to question.

Recent numerical simulations have suggested that condensed objects should be
ejected at high velocity before they have ``finished" forming in a cluster.
Prior to this work, a handful of pre-main-sequence stars have been shown to
be moving at speeds $> 10$ \kms.  To the best of our knowledge, though, the
analysis of PV Ceph and NGC 7023 described here is the first observational
work associating a speeding young star with a distant ancestral cluster.
These high-speed ejections from clusters will create a class of
rapidly-moving young stars in molecular clouds.  If these ejections are at
all common, their existence confounds both calculations of clouds'
star-forming efficiency and theories of star formation that do not allow for
stars to move rapidly through a reservoir of star-forming material while
they form.

\end{abstract}


\section{Introduction}

The ``knots" evident in HH flows from young stars come in pairs located at
similar
(but not necessarily exactly equal) distances from the powering star.  It is
ordinarily presumed, and we agree, that diametrically opposed knots at
roughly equal
distance from the star are created together.   Each knot shows the position
of a shock
caused by the interaction of ejecta from the young star with its
environment.  Much
evidence ---including the prevalence of multiple knot pairs in HH flows---
points to the
flows being highly episodic \citep [e.g.,] []{2001ApJ...551L.171A,
2001ARAandA..39..403R}.  The
episodicity is probably the result of sporadic accretion onto the young
star. 
Thus, as Bo Reipurth has emphasized, the knots represent a ``fossil record"
of a young
star's accretion history. In this paper, we use this fossil record in a new
way ---to
track the motion of the young star.  Our principal result is the mildly
dramatic
suggestion that the Young Star PV Cephei is moving at roughly 20 \kms\
through the
interstellar medium.

PV Ceph is a Herbig Ae/Be star \citep {1994ApJ...433..199L} located about
500 pc from the Sun \citep{1981ApJ...245..920C}.   The ``giant" Herbig-Haro
flow from PV Ceph, which extends for more than a pc on the sky, was
first mapped optically by \citet {1997AJ....114.2708R} and \citet
{1997AJ....114..265G}, and the molecular outflow associated with that flow
has been
extensively studied by \citet{2002ApJ...575..911A, 2002ApJ...575..928A}.
The source is a favorite of amateur optical astronomers,
because the nebulosity (known  as GM29 or ``Gyulbudagyan's Nebula")
associated with it  \citep [] [see  WFPC2 image in Figure
1]{1977PAZh....3..113G} is variable on
the time scale of months \citep {1981ApJ...245..920C}.  While extensive
observations of PV Ceph and its outflow are
available in the literature\footnote {Arce \& Goodman 2002a provides a
lengthy introduction to this
source, which we will not repeat here.}, there are, to the best of our
knowledge, no prior measurements of the
source's proper motion, or claims, as we make below, that it is moving
unusually fast.  

Analysis of PV Ceph's environs leads us to believe that PV Ceph was ejected
from the young cluster NGC 7023.   We close this paper with a brief
discussion, based on both theoretical and observational results, of how many
young stars may be ``speeding"  far from their ancestral homes.

\section{Evidence for Source Motion}
\label{evidence}
In United States courts, the ``beyond a reasonable doubt" standard is used
in criminal cases, but for
civil disputes, another standard, known as ``preponderance of evidence" sets
the bar.  We should state at
the outset that we can convict PV Ceph of speeding (at roughly 20 \kms) on a
``preponderance" of
circumstantial evidence. However, as there are no direct proper motion
measurements of PV Ceph\footnote{We have placed an upper limit on the motion
of PV Ceph by comparing the Palomar Plates with the Digital Sky Survey
(which gives a time baseline of about 43 yrs).
That limit, discussed further in section \ref {speeders}, is 33 \kms, in the
plane-of-the-sky.} we
cannot in good conscience offer a conviction ``beyond a reasonable doubt."
Below, we will go into detail
on the following three pieces of damning evidence, listed here in rough
order of decreasing importance:
\begin{itemize}
\item Observed positions of HH knots on parsec scales are better explained
in a scenario where PV Ceph is moving at a constant velocity (of $\sim 20$
km s$^{-1}$) than they are by hypothesizing angular separations of jet and
counterjet of an angle other than $180\arcdeg$ or by allowing for unpaired
knots. 
\item Redshifted molecular gas appears to ``trail"  behind PV Ceph to the
East, in addition to forming a jet along the HH flow.
\item Modeling the tilted ``wiggle" of the southern jet-like molecular outflow
implies source motion, at $\sim 20$ \kms, if the  current outflow axis is nearly
N-S, as implied by HST images.
\end{itemize}
Below, we explain this evidence, and then discuss the plausibility and
implications of PV Ceph, and possibly other young
stars, speeding through the ISM at velocities $\sim 20$ \kms.
\subsection{Geometry of the Knots in the HH flow}
\label{geometry}
We were first alerted to the possibility of a high ``peculiar" velocity for
PV Ceph
when we were trying to understand the relation between its optically
detected giant
Herbig-Haro flow \citep{1997AJ....114.2708R} and the properties of its
molecular outflow
(Arce \& Goodman 2002a,b, hereafter AGa and AGb, respectively).
 In trying to create
``pairs" amongst the HH
knots in the PV Ceph flow (see Figure 1), we initially saw two
possibilities.  First, we tried drawing straight lines {\it through} PV Ceph
and finding paired HH knots along those lines on either side of the star.
Attempting that, we noticed that several knots would be without partners
on the opposite side of the source.  No obvious explanations (e.g. high
extinction) for the missing knots were very relevant.  Second, we tried
to find pairs of knots at similar distances from the source, the method
of pairing one might expect from a flow emitting ``ballistic" knots.
Doing this, we noticed that lines drawn between pairs of
knots at very similar distances did not all pass through the source.
Instead, the halfway (a.k.a. bisector) point only passed through the
source for the innermost pair of knots, and the halfway points for
progressively more distant knots
seemed to be moving along a line to the East (see Figure 1).  The simplest
explanation for this motion of the
bisection points is that the outermost HH knot pair is oldest, and that the
source has been
moving, at roughly 20
\kms\ (see below) to the West, at least since the initial
ejection.\footnote{An alternative explanation is
that every jet emitted from (an almost-stationary) star bent in such as way
as to give the current distribution of
HH knots, where each pair is then separated by an angle significantly $\neq
180\arcdeg$.  We consider this explanation too much of a conspiracy to be
plausible.} In the moving-source scenario we favor, the position angle of
the flow is
also changing, appearing to approach a N-S direction in a counter-clockwise
direction (see Table 1).

\subsubsection{The Plasmon Model}
\label{plasmonmodel}

Imagine that a star moves in a straight line and periodically emits blobs of
high-velocity gas that lead to HH knots.  Assuming that the star's motion is
very slow compared with the initial knot (jet) speed, the
key factor determining the path of a knot once
it is emitted from the moving star is how quickly it decelerates in
the ambient medium.\footnote{We are very grateful to James Yorke for asking
a
question 
which caused us to fully appreciate this point.}  In the limit of no
deceleration (a zero-density ambient
medium), the component of a knot's velocity along the star's direction of
motion will equal that of the star.  So, even as a pair of knots moves
outward from the star, the bisector of a line between them will never be
seen to ``lag" behind the star.  In other words, with no deceleration by an
ambient
medium, one could never tell that the source of an HH-flow was moving from
the knot pattern,
because everything would apparently take place in a rest frame defined by
the
star.   In the opposite limit of a very high-pressure ambient medium, a knot
will rapidly lose the component of its velocity created by the star's
motion:  as it moves outward in the ``jet" direction, it will lag behind the
source by an amount roughly proportional to the distance moved ``outward" by
the knot scaled by the ratio of star-to-knot velocity (see below).\footnote{It is interesting to realize here that a model of a source moving through a medium at rest is quite equivalent to that of a stationary source in a moving medium.  We thank the referee for pointing us to the elegant work on ``HH jets in supersonic sidewinds" by Cant\'o \& Raga (1995), Lim \& Raga (1998) and Masciadri \& Raga(2001).  The estimates of PV Ceph's velocity one can make using the formalism in Cant\'o \& Raga (1995) is in line with the model we offer here.}

For giant HH flows, on pc scales, the velocities, pressures and densities
involved are such that neither the ``zero-density" ambient medium
(ballistic) or the ``very rapid deceleration" picture typically applies.
Instead, the
deceleration history of the knots is critical to the geometry observed.

In 2000, Cabrit \& Raga  successfully modeled knot
deceleration in the HH 34 ``superjet" by assuming the source to be
stationary
and  using what is
conventionally known as a ``plasmon" model (first studied analytically by De
Young \& Axford 1967; see also Cant\'o et al. 1998).  In their Figure 1,
Cabrit \&
Raga show that
a plasmon model where clumps are decelerated simply by their interaction
with a uniform
density medium produces an excellent fit to the observed knot velocities in
HH
34.  

In this paper, we modify the plasmon model used by Cabrit \&
Raga (2000, in turn based on Cant\'o et al. 1998), to calculate the position
of HH knots in a flow from a {\it moving}
source, as a function of time.  To do this, we treat each
knot as a blob of gas, decelerating as it travels through an ambient medium,
and
we assume that blobs are ejected episodically, in diametrically-opposed
pairs,
from a source moving at constant velocity.

To model the knots in the PV Ceph flow as a set of decelerating ``plasmon"
blobs created by an episodic jet,  we use the following generalizable
procedure.  Assume that a jet is emitted in a direction defined in space by
$\hat{v}_j$, from
a source which is moving in a straight line defined by the
direction $\hat{v}_{*}$.  These two directions define the $v_j-{v}_{*}$, or
the ``jet-source motion,"
plane, and the angle between them is
$\theta_\circ$.  The path of the knots in the $v_j-{v}_{*}$ plane is then
readily modeled by numerically integrating
\begin{equation}
\label{voft}
\vec v(t)=\int {{d\vec v} \over {dt} }dt + \vec v_{j} + \vec v_{*}.\,
\end{equation}
where the deceleration of the plasmon, $d \vec v/dt$, is calculated using
eq. \ref{dvdt}, below.  To compare this
model with observations, one can either re-calculate the motion as projected
from the  $v_j-{v}_{*}$ plane to the plane-of-the-sky frame, or
de-project the observed positions and velocities into the more natural
$v_j-{v}_{*}$ plane itself.  We choose to do the latter, as our model is
then more
intuitively understandable.

For a hypersonic, isothermal clump decelerated only by the gas drag of the
ambient medium, the instantaneous
deceleration can be written:
\begin{equation}
\label{dvdt}
{{dv} \over {dt}}=-\left( {{{\xi {\rho_{\rm amb}}c^{4} \over M}}}
\right)^{1/3}v^{2/3},\end{equation}
where $v$ is the clump velocity at any instant, $c$ is (traditionally) the
isothermal sound speed, $M$ is the constant mass of the clump, $\rho_{amb}$
is the mass density of the ambient medium, and $\xi=14$  is a constant
describing the shape of the plasmon (adopted from Cant\' o et al. 1998).
For our purposes, we take $c$ to be the typical line width in the ambient
medium, since that describes the true ambient pressure, which far exceeds
the isothermal sound speed.

While the instantaneous value of $dv/dt$ is itself a scalar, we use $d \vec
v/dt$ in eq. \ref{voft} to mean the deceleration along the direction of the
jet at a given time, $t$.  In the numerical integration described above, we
update the direction of $\vec v$ for each time step in the calculation.
This kind of numerical directional updating is obviously unnecessary for a
stationary source, where the velocity of the plasmon  changes only in
magnitude, and not in direction\footnote{In a reference frame fixed to the sky, the path of each knot in a knot pair is a straight line (see Figure 7).  It is only in a frame tied to the source that a knot's path appears to curve.  Figure 2, discussed later in this section, shows the locus of all knot positions in a jet-source plane where the jet direction is fixed and (0,0) corresponds to the source position.}.
In that stationary-source case, the velocity of the knot as a function of
radius can be written analytically, as
\begin{eqnarray}
\label{vblob}
{{v_{knot}(r)} \over {v_j}}=\left( {1-{r \over {d_\circ }}} \right)^{3/4}\\
\label{dcirc}
d_\circ ={3 \over 4}\left( {{{Mv_j^4} \over {\xi \rho _{\rm amb} c^4}}}
\right), 
\end{eqnarray}
where $r$ is measured along the star-knot direction (Cant\'o et al. 1998,
Cabrit \& Raga 2000).

Unlike the situation in HH34 where Cabrit \& Raga could fit their plasmon
model to detailed observations of knot velocities, there is no extensive set
of velocity measurements for the PV Ceph flow.   Instead, David Devine's
Ph.D. thesis gives radial velocity measurements of for just three
knots:\footnote{Note that due to historical (and nomenclature) reasons, the
``PV Ceph"
flow is comprised of two sets of HH objects, with different numbers.
``HH 215" refers only to the jet very close to the source, and ``HH 315"
includes all the knot pairs we are modeling here.  See Figure 1 for a
finding chart.} $v_{\rm radial}\sim -271$ \kms\ for HH215A, and $v_{\rm
radial}\sim -55$ and 52 \kms\ for HH315B and HH315E, respectively \citep
{Devine97}.  What we {\it do} have in PV Ceph are detailed measurements of
current knot {\it positions} (Table 1), and very good observational
estimates of the ambient velocity dispersion and density (Table 2).   We
only require that our moving-source plasmon model fit the observed knot {\it
positions} (Figure 2).    We use the few velocity measurements available as
a way to: 1) estimate a velocity for PV Ceph's jet very close to the source;
and 2) check that the velocities predicted by our model for knots farther
from the source are reasonable.

To estimate the jet velocity close to the source, we
conservatively assume an inclination angle of the HH 215 flow of 45\degrees\
(as in Devine 1997).  The measured radial velocity for HH215A then implies
that the initial 3D velocity of ``a"  jet from PV Ceph  is of order nearly
400 \kms.   The same 45\degrees\ inclination angle implies that our plasmon
model should predict velocities  of order 100 \kms\ for the ``middle" pair
of knots (B-E) in the HH315 flow.  Keep in mind, however, that there is no
reason to assume that the initial ejection speed ---or the inclination
angle--- of the jet(s) that created each pair of knots in the PV Ceph flow was
the same ---this is only a simplifying assumption in our ``fitting" of a
plasmon model.  In reality, it is likely that the jet is episodic, emitting
bursts of varying strength \citep {2001ApJ...551L.171A}, and that its
inclination angle varies ---especially given that we know its position angle
on the plane of the sky varies.  Nonetheless, in order to make a simple
model of knot deceleration, we ignore the second-order effects that
variability in ejection velocity or $\theta_{\circ}$ have on the plasmon
calculation.

The dark curve (labeled ``22 \kms") in Figure 2\footnote{The  (0,0) of the
coordinate system in Figure 2 is tied to the position of
PV Ceph.} shows the locus of possible knot positions in the $v_j-{v}_{*}$
plane for the moving-source plasmon model outlined above, under the physical
conditions listed in Table 2.
Exactly where knots lie along this curve depends only on when they were
emitted.  To facilitate data-model comparison, the three black filled
circles show positions corresponding to the PV Ceph knots ---if the change in jet ejection angle
with time is ignored\footnote{Figure 7 shows an illustration of what the same (22 \kms) model shown in Figure 2 gives when the jet angles are purposely adjusted to match those listed in Table 1.}.  As the figure shows, the observed knot positions
in the PV Ceph flow are remarkably well-fit by a simple 22 \kms\ plasmon
model. The two lighter curves in Figure 2 show how sensitive the model is to
assumptions about the velocity of the central source.  As expected, for a
source velocity very slow ($v_{*}=1$ \kms) compared with the jet velocity
($v_{j}= 350$ \kms), the knots would not be seen to lag behind the star at
all, whereas for a speeding source ($v_{*}=100$ \kms), the knots would be
seen to lag almost immediately.

\subsubsection {Assumptions and Uncertainties in the Plasmon Model}
While the ``22 \kms" source velocity curve shown in Figure 2 and
parameterized in Table 2 clearly gives a very good fit to the PV Ceph flow's
knot positions, we are {\it not} claiming that 22 \kms\ is the ``exact"
velocity of the source.  Instead, after an extensive search of parameter
space in our plasmon model, we find the set of parameters listed in Table 2
to come closest to fitting our data.

We accomplish the seemingly impossible trick of fitting six data points
(two-dimensional positions for 3 pair of knots) with a model that requires
nine inputs by constraining many of the inputs with additional observations.
As shown in Table 2, the density and velocity dispersion in the calculated
plasmon model are tied to observations of PV Ceph's host cloud in \co\ and
\thco\ (AGa).   The angle between $\hat
v_{j}$ and $\hat v_{*}$ is taken to be 90\degrees\ because the current $\hat
v_{j}$ is clearly N-S (see WFPC2 image in Figure 1) and $\hat v_{*}$ appears
to be from E to W based on the geometry shown in Figure 1.  The inclination
angle of the flow to the sky is assumed to be 45\degrees, as a conservative
guess\footnote{\citet {1997AJ....114..265G} model the HH 315 flow as
emanating from a stationary precessing source, and derive an inclination
angle of just 10\degrees\ to the plane of the sky.   Since our current
analysis strongly suggests that the HH knot positions in the flow are
critically effected by source motion (and are not as well-modeled as arising
from a stationary precessing source) we do not adopt the 10\degrees\
inclination angle the G\'omez et al. model implies.}, and the inclination
angle of the star's motion w.r.t. the plane-of-the sky is assumed to be
20\degrees, since any much larger angle would not give PV Ceph sufficient
plane-of-the-sky motion for us to have noticed it.  The mass of each ejected
blob, $3 \times 10^{-4}$ M$_\odot$, and the geometric factor describing the
shape of the plasmon, $\xi=14$, are assumed to be similar to the values
Cabrit \& Raga found for HH 34.   So, with those seven parameters  held fixed
at the values shown in Table 2, we can ``fit" the positions of the HH knots
in the PV Ceph flow by {\it only adjusting the star and initial jet
velocities}.  If we use the radial velocity for HH 215A measured by Devine
(see above) to set  $v_{j}\approx 350 \rm {\ to\ } 400$ \kms, we find
$v_{*}\approx 22$ \kms\ as a ``best fit."

After stretching each of the input parameters to within what seem reasonable
bounds based both on our observing experience and on where they cause the
plasmon model to grossly fail, we are willing to predict, that the ``true"
three-dimensional velocity of PV Ceph is more than 10 but less than 40 \kms.

\subsubsection{Utility of Plasmon Models in the Real ISM}

In reality, the velocity and mass of each knot emitted in a giant HH flow is
surely not the same \citep{2001ApJ...551L.171A}, and the density and
velocity dispersion into which the knots travel is not uniform.  Thus,
despite what may appear as exquisite model-data agreement in Figure 2, the
plasmon model should not be taken as more than a 1st-order estimate of the
interaction of HH knots and the ambient ISM.

In the lower panel of Figure 3, we show knot velocity as a function of
distance from the source, for the ``22 \kms" moving-source plasmon model.
Grey circles are plotted ---on the calculated curves--- at the deprojected
distances of the observed knot pairs.  As mentioned above, in the HH 315
flow, Devine (1997) measured radial velocities only for knots B and E, both
of which deproject to about 75 \kms, assuming a 45\degrees\ inclination
angle for the flow.   The strict plasmon model shown in Figure 3 would
predict twice that velocity for the B-E knot pair.  Does this mean our model
is wrong?  No ---it could well mean that ignoring changes in jet ejection angle in PV Ceph
would not be a good assumption in a more detailed model. If the jet giving
the B-E knot pair is more along the line-of-sight than the others, then the
knots with what seem ``too low" velocities are in fact further from the
source in 3D, as the model predicts for the observed radial velocities (see
Figure 3). Alternatively, if the B-E jet is more in the plane-of-the sky
than the others, Devine's measurement of its radial velocity would need to be
``deprojected" to 3D by a larger correction factor, again bringing it into
line with the model.

Another explanation altogether for any detailed mismatches between model and
data is simply that the velocities with which knot pairs are emitted in the
PV Ceph flow is variable enough to explain the discrepancies.  If we had
detailed measurements of the radial velocities and proper motions of all of
the knots in PV Ceph, we could construct a much more sophisticated model of
the flow's history and geometry.  The plasmon ideas explored here will lie
at the heart of future more sophisticated models, and as such, they are very
useful.

Figure 3 also makes another important point ---about the age of outflows.  The
top panel of Figure 3 shows that if knots slow down at the rate implied by
the plasmon model, then estimating the age of an outflow by dividing the
distance of the outermost knot by the knot's observed speed (i.e.
calculating a dynamical time) will significantly overestimate the age of the
flow.  The plasmon model of PV Ceph's flow implies that the slowest
currently-supersonic blob (velocity marginally greater than the ambient
dispersion of 3.2 \kms) would have left PV Ceph 10,000 years ago.  For
comparison, the dynamical age of PV Ceph implied for a knot moving at this
minimum speed (3.2 \kms) would be 560,000 years!  In reality, velocities of
several tens of \kms\ are required to produce shocked gas visible optically,
and the dynamical time of the ``outermost" knots in a giant Herbig-Haro flow
are likely to overestimate the age of the flow by about an order of
magnitude (see Figure 3).

The bottom panel of Figure 3 shows that blobs will slow to velocities $< 10$
\kms\ at $\sim 2$ pc in the PV Ceph flow, and this slowdown ---which limits
the observable size of a flow--- will happen at some similar radius in other
flows at well.\footnote{In the two HH flows where direct proper motion and
radial velocity measurements are both available [HH34 \citep
{1997AJ....114.2095D}, discussed in \S \ref{plasmonmodel}, and HH7-11
\citep{2001AJ....122.3317N}] the data show decreasing knot velocity with
source distance.}  

Thus, taking the results of the two panels of Figure 3 together, we see that
the ``size" of an outflow we see now is determined only by how far the
oldest knot {\it we can still see as moving and shocked} has traveled in to
the ISM, and estimating the age of the flow based on the parameters of this
knot is a flawed procedure.  The flow's age will be overestimated by the
dynamical time of the outermost knot ---and, perhaps worse, the age will be
underestimated by assuming that outermost knots we see date the flow's
birth.  Older, unseen, knot pairs may have blended back into the ISM by now.
Thus, the plasmon analysis leaves us realizing that many outflow ``ages"
estimated in the extant literature may be incorrect.

\subsection{A Trail of Redshifted Gas}
\label{trailsec}
Position-velocity diagrams of \cotwoone\ and $^{13}$CO(1-0) emission each show a
very peculiar extension to highly-redshifted velocities right at the
position of PV Ceph (see Figure 15 of AGb, and Figure 4 of AGa, respectively).
If PV Ceph is indeed zooming
through the ISM with a velocity near 20 \kms, then the moving blob of gas
gravitationally bound to PV Ceph is likely to leave a trail of accelerated
gas in its wake.   The trail might be created in one of two ways.  First,
one could expect gas to be ``swept-up" by the action of the dense,
supersonic, blob in much the same way it is in the ``prompt-entrainment"
picture of jet-driven outflow.  Alternatively, if the pressure drop created
in space after the moving blob has passed-through is great enough, molecular
gas clouds could condense out of lower-density warmer ambient material.
This second way to form a trail is analogous to the formation of ``vapor" or
``con" trails behind airplane wings.

In the two short sections below, we first discuss the observations of the
trail of mildly redshifted molecular gas (\ref{trail}) created by PV Ceph's
motion, and then consider how much gas is likely to be bound to PV Ceph
(\ref{groupies}).  In this paper, we do not consider in detail whether the
swept-up trail or ``con-trail" scenario is more likely, but we suggest this
as an interesting question for future work.

\subsubsection{Trail in the 100 \cc\ Molecular Gas}
\label{trail}

Detailed mapping of the \cotwoone\ emission, which traces densities $\sim
100$ \cc, around PV Ceph has revealed a ``trail" in the mildly redshifted
gas distribution in exactly the direction PV Ceph would have come from in
the model described in \S\ref{geometry}.

The  contour map labeled ``more redshifted" in Figure 1 shows \cotwoone\
emission integrated from 4.15 to 6.46 \kms\ (based on Figure 6 of AGb).
This highly redshifted molecular gas is
associated with the current jet emanating from PV Ceph, whose
plane-of-the-sky orientation (like the optical flow in the Figure 1 panel
labeled ``WFPC2") is almost exactly N-S (\S \ref{wiggle}).  The ``ambient"
\cotwoone\ LSR velocity 
near PV Ceph is $\sim 2.5$ \kms\ (AGa).  The panel
labeled ``less redshifted" in Figure 1 shows \cotwoone\ emission integrated
from 3.16 to 4.15 \kms\ (based on Figure 6 of AGb), and it shows two
important features.  First, the jet itself is obviously still present,
although its orientation, especially near the tip, curves back toward the
East.  That curvature is consistent both with our proposed general clockwise
rotation of the jet direction over time, and/or with motion of the source
from East to West.  (Slower material is older in the
decelerating-jet-from-a-moving-source picture and would lie to the East of
the faster material.)  Second, and more importantly, the ``less redshifted"
map shows a {\it trail} in the molecular gas, extending all the way to the
edge of this map, in exactly the direction PV Ceph would have come from in
the moving source model.  In fact, the same kind of extension is evident on
larger scales.  The pc-scale (main) image of the outflow in Figure 1 also
clearly shows an extension of redshifted emission toward the East.  For
readers unfamiliar with outflow maps,  note that this kind of asymmetry is
{\it not} common in these maps.

We propose that the ``redshifted" trail of emission seen in \cotwoone\ to
the East of PV Ceph is caused by the interaction of the ambient medium with
a gravitationally bound  higher-density blob (see \S\ref{groupies})
traveling along with PV Ceph,  moving E-W and away from us at an angle of
$\phi_{\rm star-sky}=20^{\circ}$ (see Table 2).  Motion away from us would
redshift the ``trailing" material with respect to the cloud velocity.
  If we assume that the full E-W stretch of the redshifted
material in the main panel of Figure 1 (at the declination of PV Ceph) is
due to gas created by this interaction, PV Ceph would now have travelled at
least 0.2 pc.  At 20 \kms, traveling 0.2 pc takes 10,000 years, which is
then a lower limit on the age of PV Ceph.

\subsubsection{Dense Gas Moving With PV Ceph}
\label{groupies}
The upper left panel of Figure 1 shows a map of \thco\ emission between 3.15
and 3.71 \kms\ near PV Ceph (from AGb, Figure 7).  In other words, this is a
map of $\sim 1000$ \cc\ gas moving with an average velocity redshifted by
about 1 \kms\ with respect to the ambient material in the vicinity of PV
Ceph. Since the gas traced by \thco\ is only $\sim 10\times$ denser than
that traced by \cotwoone, we expect that most of the redshifted \thco\
evident in the upper left panel of Figure 1 was moved to these velocities by
the same mechanism that redshifted the \co\ discussed above in
\S\ref{trail}.   It is possible, though, that \thco\ is a high-enough
density tracer that we could detect a {\it small} clump gravitationally
bound to PV Ceph moving along with it.    The size of that clump depends on
a competition between gravity, which acts instantaneously to draw material
into a blob around a point source, and gas drag, which tries to keep
material where it was, but can only do so at a speed governed by the
pressure of the gas.

To illustrate how much gaseous material might move along with a speeding
protostar, Figure 4 shows the modified ``Bondi-Hoyle" accretion radius for a
point mass moving through an ambient medium (see Bonnell et al. 2001),
\begin{equation}
\label{BH}
R_{BH}={2GM_{*} \over {\sigma^{2}+v_{rel}^{2}}}={GM_{*} \over
{\sigma_{eff}^{2}}},
\end{equation}
where $v_{rel}$ is the dimensionally-averaged speed with which the point
mass ($M_{*}$) moves through the ambient medium whose one-dimensional
velocity dispersion is $\sigma$.
For PV Ceph, using Bonnell et al.'s formalism, $v_{rel}=7.3$ \kms, the
average of 22, 0, and 0 \kms.  The velocity dispersion around PV Ceph is
much less than this, so we can ignore the $\sigma^{2}$ term in eq. \ref{BH},
making $\sigma_{eff}\approx v_{rel}/\sqrt 2\approx 5.2$ \kms.  The mass of
PV Ceph is not precisely known, but is likely to be $< 7$ M$_{\odot}$ (less
than the most massive Ae/Be stars).  So, examining Figure 4 for the
conditions relevant to PV Ceph, we see that a bound ``clump" moving along
with a 7 M$_{\odot}$ star would have a diameter of $<0.001$ pc, or about 0.5
arcsec at 500 pc.  Thus, even the highest-resolution molecular-line
observations shown in Figure 1 (\thco\ map at the upper left) could not
spatially resolve the ``Bondi-Hoyle" clump that would be speeding along at
$\sim 22$ \kms\ with PV Ceph.

In our FCRAO observations of PV Ceph, published in AGa, we also searched for
N$_{2}$H$^{+}$ emission associated with the star.  Ordinarily,
N$_{2}$H$^{+}$ would reveal a $> 10^{4}$ \cc\ ``dense core" associated with
such a powerful outflow source.  Curiously, no N$_{2}$H$^{+}$ emission was
detected.  We suspect that PV Ceph's rapid motion, which would make a bound
core very small (see Figure 4) is responsible for this state of affairs, and
we discuss how a star might have an outflow with no dense core in \S
\ref{nagging}.

In principle, Bondi-Hoyle accretion is not the complete mechanism
determining how big a blob would {\it appear} to surround a rapidly-moving
young star.   While it is true that PV Ceph itself can only carry along a
small ($\sim 0.001$ pc) blob, that very dense blob will in turn drag less
dense gas along with it, at a lower speed.  The asymptotic evolution of this
process creates the appearance of a larger blob when observed at a density
lower than that of the clump ``bound" to PV Ceph.  The gas physics
associated with this process are not dissimilar to those of the plasmon
problem discussed in \S\ref{plasmonmodel}, but have not, to our knowledge,
been exactly worked out for the specific case of a gravitating
point-mass$+$cocoon moving through a relatively high-pressure medium. Thus,
we conclude this section with the suggestion that the morphology of the
mildly-redshifted \thco\ emission around PV Ceph might be well-explained by
a detailed gas-dynamical model of a point-source plus it's
gravitationally-bound/swept-up cocoon moving through a molecular cloud.

\subsection{Tilted Wiggle in the Innermost Jet}
\label{wiggle}
The  wiggles in the jet of ``more-redshifted" gas shown in the lower right
corner of Figure 1 (see AGb, Figure 18, for detail) are
well fit by a model where they arise from
a precessing jet, the source of which is moving at roughly 20 \kms.

The inset HST WFPC2 image (Padgett et al.~1999, and Padgett, personal
communication) of the inner region of the PV Ceph outflow shown in Figure 1
shows the current symmetry axis of the PV Ceph biconical nebula to be very
close to N-S. Thus, we take a position angle of zero
degrees, shown as a yellow line in the lower panels of Figure 1, to be the
orientation of the jet coming from PV Ceph ``now".  This orientation is
consistent with the counter-clockwise rotation of PV Ceph's flow implied by
the arrangement of large-scale HH knots (see
Figure 1 and Table 1), and with the analysis of the redshifted molecular jet
presented in \S\ref{trail}.

Figure 5 shows the Right Ascension (column number) of the peak emission in
the ``more redshifted" \cotwoone\ gas (shown in the lower-right panel) as a
function of declination (row number; see AGb, Figure 18).
An excellent fit to these data is obtained by modeling them as the sum of a
constant source motion (shown here as the sloping line in the left panel
Figure 5) plus a precession of the jet (sine wave shown in the right panel
of Figure 5).  

As we stated in \S\ref{plasmonmodel}, in the limit of a very high-pressure
ambient medium, a knot will rapidly lose the component of its velocity
created by the star's motion:  as it moves outward in the ``jet" direction,
it will lag behind the source by roughly the distance moved outward by the
knot times the ratio of star-to-knot velocity. This ``high-pressure" limit
is relevant to the redshifted CO jet-like lobe modeled in Figure 5
(since it lies in a high-density region), so the slope of
the linear fit shown in the left panel implies that
\begin{equation}
\label{vwiggle}
v_{*}=0.06 \times v_{jet}.
\end{equation}
For the {\it same} 350 \kms\ jet velocity assumed in \S\ref{plasmonmodel}
(see Table 2), equation \ref{vwiggle} gives $v_{*}=21$ \kms \/ ----in
astonishingly good agreement with the independently-derived source velocity
based on the plasmon model!\footnote{Possible origins of the precession
implied by the sinusoidal variations are
discussed in detail in AGb.}

If instead of assuming that the jet is now perfectly N-S,  we instead assume
an orientation more toward the overall tilt of the apparent \cotwoone\ jet,
the source velocity is reduced.  But, keep in mind
that the HST image essentially rules out that the current jet direction is
the one given by even the most redshifted \cotwoone\
gas (Figure 1, bottom-right panel).  Alternatively, if we increase our
estimate of the jet velocity then the source velocity will increase.  The
key point remains, though, that for the most reasonable assumptions, of a
N-S current jet orientation, and a 350 \kms\ jet velocity, our modeling of
the wiggles in the south jet-like redshifted lobe implies the {\it same $\sim 20$ \kms\ velocity
for PV Ceph} as the plasmon model of the larger scale (HH 315) flow.

\section{Possible Causes of ``Speeding"}
\label {causes}

If PV Ceph really is moving at $\sim 20$ \kms, how did it get such a high
velocity? Below we consider several hypothetical scenarios, in order of
decreasing {\it a priori} likelihood.

\subsection{Runaway star from a young stellar cluster}
\label{wow}
The nearest (known) stellar cluster to PV Ceph is NGC 7023, a cluster with a  stellar mass of roughly 100 M$_{\odot}$ \citep{1969AJ.....74.1021A} whose most massive member is the B3e star HD200775 with mass 10 M$_{\odot}$ \citep {1958ApJ...128..207M,1972ApJ...173..353S,1978ApJ...220..510E}.  Figure 6, which
includes a large-scale 100 $\mu$m IRAS image, shows NGC 7023 to lie in {\it
exactly} the direction PV Ceph would have come from according to our
analysis in \S 2 (see also Figure 1).  The LSR velocities of the molecular
gas associated with NGC 7023 are primarily between 2 and 3 \kms, as is the
case for PV Ceph \citep {1998ApJ...500..329G}. The Hipparcos distance to NGC
7023 is  about 430 pc \citep {1997A&A...323L..49P}.  If PV Ceph is at the
same distance (which is consistent with uncertainties), then it is currently
about 10~pc away from the center of NGC 7023.

We emphasize that the long arrow shown in Figure 6\footnote{The slight
curvature of this arrow accounts for the map projection scheme, and in fact
represents a straight-line path through space.} is just the extension of the
short line segment shown as the path of PV Ceph in Figure 1.  We were very
surprised to find that path pointing straight at NGC 7023.  The probability
of the position angle of PV Ceph's path on the plane of the sky accidentally
matching a PV Ceph-NGC 7023 connector to better than a few degrees is only
about 1 percent (3\arcdeg /360\arcdeg).  Combining the similar distances of
NGC 7023 and PV Ceph with the low probability of randomly finding the
``right" plane-of-the-sky orientation for PV Ceph's path, we know that at
least three of the four space-time dimensions strongly imply that PV Ceph
came from NGC 7023.

What about the time dimension?  At its current velocity of $\sim
20$~km~s$^{-1}$ it would have taken PV Ceph $\sim 5 \times 10^5$~yr to reach
its current position. This amount of time is approximately  equal to or
greater than the estimated age of PV Ceph ---which is thought to be a Class I
protostar of spectral type A5 with an age $\sim 10^5$~yr (Fuente et
al.~1998)\footnote{This age estimate assumes a ``normal" past for PV Ceph.
If, in fact, it spent some large fraction of its life wandering at high
speed across space, age estimates may need revision!}.  On the other hand,
if PV Ceph came from NGC 7023, its {\it initial} velocity could have been
larger.\footnote{We thank Laurent Loinard for inspiring us to
consider this idea.} Without knowing where in the gravitational potential of
the cluster PV Ceph might have originated, it is very difficult to estimate
its potential initial speed, but it is reasonable to guess $> 20$ \kms. For reference, to acquire a velocity of 22 \kms\ by converting gravitational to kinetic energy in a ``slingshot" encounter with a star of mass $M_{scat}$ requires a closest approach of $R_{c}$[A.U.]$=4.5 M_{scat}/M_{\odot}$ (e.g. 45 A.U. for a 10 M$_{\odot}$ star).
N-body simulations of open star clusters show that dynamically ejected stars
can reach velocities of up to 200 km~s$^{-1}$ (e.g., Leonard \& Duncan 1990,
see also de La Fuente 1997, and \S 3.2, below). If PV Ceph was in fact
ejected at $> 20$ \kms, NGC 7023's gravity would not do much to slow the
star and whatever disk or blob was bound to it, but the cumulative effect of
all the matter (shown as green and yellow in the dust emission map in Figure
6) between NGC 7023 and PV Ceph's current position could potentially have
slowed it ---through primarily drag forces--- to 20 \kms.  Thus,
the $\sim 5 \times 10^5$~yr to get from NGC 7023 to PV Ceph's current
position is an upper limit, and it is plausible that PV Ceph was born in
NGC 7023.

The inset Sloan Digital Sky Survey Image of NGC 7023 in Figure 6 reveals a
striking dark swath across the Southern part of the cluster.  The Figure also
shows that the 174\arcdeg E of N path of PV Ceph implied by our analysis of
its outflow (\S 2) would have it coming from a direction nearly perfectly
aligned with this dark lane.  It is surely tempting to think that PV Ceph's
ejection from NGC 7023 might have caused this ``exit wound."  The lane is,
presently, of order 0.1 pc wide.  The ambient (\thco-traced molecular) gas in NGC 7023 has a
velocity dispersion  $\sim 1.5$ \kms\ \citep {2003AJ....126..286R}, which
implies that closing the current gap by turbulent diffusion would take about
$8 \times 10^{4}$ years, that is 6 times less than the upper-limit for the
time since PV Ceph's ejection.   Therefore, if the rift in NGC 7023 were in
fact an exit wound caused by the escape of PV Ceph, then either the wound
must have originally been much wider, the pressure in the rift was or is higher than ambient, or PV Ceph's initial velocity was much
higher than 20 \kms.  If the rift was caused by the exit of PV Ceph from
NGC 7023, and it is now half-closed, PV Ceph's ejection would have been
133,000 years ago, and its time-averaged speed since ejection would be almost implausibly high ---75 \kms.  Alternatively, if PV Ceph left a shock-heated trail when it created the rift we see now, the rift would have had a velocity dispersion (and pressure) greater than ambient, and would have taken longer to close.

Existing maps of the distribution of molecular gas in NGC7023 bear on the importance of the ``rift" seen at optical and infrared\footnote{2MASS and Spitzer Space Telescope \citep{Megeath2004} infrared images also show a gap in emission from NGC7023 along the direction PV Ceph would have exited.} wavelengths in two ways.  First, HD200775 (the 10 M$_{\odot}$ star that illuminates NGC7023) is centered in a  bipolar cavity which is oriented roughly E-W  \citep {1986A&A...163..194W,1998A&A...339..575F, 2003AJ....126..286R}, so that the apparent rift does lie within, or possibly near the edge of, the western side of the cavity. We note, though, that the opening angle of the cavity is much larger than that of the rift seen optically.  Second, the integrated emission from \co\ in NGC7023 shows a very distinct hole \citep{1978ApJ...220..510E} in what appears to be the place where PV Ceph would have punched out of the cloud, if it indeed had traveled down the optical ``rift" and out the far side of the cloud (see Elmegreen \& Elmegreen 1978, Figure 1).  The hole is too large ($\sim 0.7$ pc) to have been punched by the ejection of just the star, but its existence is provacative nonetheless.

\subsection{Ejection from a Multiple Star System}

Numerical studies of the interactions of multiple star systems reveal
that most systems with
three or more members are unstable \citep
{1991ARA&A..29....9V,1998MNRAS.298..231K,1998MNRAS.301..759K}.
These studies show that within about
100 crossing times (or approximately a few times $10^4$ yr)
it is very probable that a member of the system will be
ejected (see Reipurth 2000 for details). The ejected star will not always be
the least massive member of the system.
The velocity of the
ejected member, as shown by the numerical simulations
of Sterzik \& Durisen (1995),
is  typically 3 to 4 \kms, but can reach values of 20~\kms\ and more.  In
fact, \citet {1998MNRAS.301..759K}  find escape speeds in excess of 30 \kms\
1 \% of the time in systems with between 3 and 10 stars.    The same
simulations give velocities between 10 and 20 \kms\ 20 to 30 \% of the time
and between 20 and 30 \kms\ 3 to 5 \% of the time, with the exact
percentages depending on the group's initial mass and velocity
distributions.  So, PV Ceph's $\sim 20$~\kms\ velocity would not be considered
extreme at all in the context of these simulations.

If PV Ceph obtained its current speed as a consequence of being ejected from
a  multiple star system ---and that system were something smaller than
NGC 7023--- one would expect to find the remaining members of this stellar
system ``nearby." The nearest detected sources (shown as white stars in
Figure 6) which
could be young stars are two faint IRAS point sources east of PV Ceph, IRAS
20495+6757  and IRAS 20514+6801, at 25.6\arcmin \/ and 37.4\arcmin \/ from
PV Ceph, respectively (see IRAS point source catalog). Assuming that these
two sources are at $d \sim 500$~pc, then they are at distances of  3.7~pc
and 5.4~pc from PV Ceph. At a velocity of about 20~km~s$^{-1}$ it would have
taken PV Ceph $2-3 \times 10^5$~yr to travel from the current
position of IRAS 20495+6757  and IRAS 20514+6801
to its current position. As was the case for the NGC 7023 ejection scenario,
this time is a good fraction of the estimated age of PV Ceph. Even if the
different members of the putative
multiple system broke up half-way between the current position of PV Ceph
and these IRAS sources, it would
have taken PV Ceph about 100,000 years to reach its current position at a
speed of 20~km~s$^{-1}$.  And, again as in the NGC 7023 scenario, it is
possible that the initial ejection speed of PV Ceph was $>20$ \kms, reducing
the travel time to its current location.  Given that the back-track (shown
in purple in Figure 6) of PV Ceph aligns so perfectly with NGC 7023, and not
very well at all with either of these IRAS sources, we find it much more
likely that PV Ceph was ejected from NGC 7023 (see \S \ref{wow}) than from a
smaller multiple system with either of these anonymous IRAS sources.

To be thorough, we mention that it is also possible that PV Ceph's
ex-companion(s)  has(have) not yet been detected, and is(are) still close
($\leq 5$\arcmin) to PV Ceph, and thus unresolved by IRAS. Reipurth, Bally,
\& Devine (1997) speculate that HH~415, near the north-western edge of the
optical nebula (see Figure~1) is formed
by another (yet undetected) young star in the region. So, while it is not
impossible that the source of HH~415 is PV Ceph ex-companion, this seems
much less likely than the NGC 7023 scenario.

\subsection{Random Motions}

Lastly, it is not crazy to think that PV Ceph's $\sim 20$~km~s$^{-1}$ speed
might be due to purely random motions. Proper motion studies of groups of
associated stars from the same star-forming region  show that the
three-dimensional velocity distribution of the studied samples have a
1-$\sigma$ dispersion about 3 to 7~km~s$^{-1}$ (e.g., Jones \& Herbig 1979;
Frink et al.~1997; Sartori et al.~2003). Thus, even though the frequency of
stars with velocities near 20~km~s$^{-1}$ (i.e., about a 3 to 5-$\sigma$
deviation from the mean) is very low, it is still a non-zero quantity.  PV
Ceph could simply be one of the few stars in the high-velocity tail of the
velocity distribution of young stars.  Finding such an outlier would not be
unlikely, since the telltale signs of motion discussed in \S\ref{evidence}
would be more apparent in a source moving unusually fast.  Again, though, we
find this possibility less attractive than the NGC 7023 idea.

\section{Implications of High Velocity in Young Stars: Admitting a More
Dynamic Star-Forming ISM}

\subsection{How Common is Speeding?}
\label{speeders}

Amongst massive main-sequence stars, speeding is a relatively common, and
oft-documented, offense \citep [e.g.][]{1991AJ....102..333S}.  Thanks to a
plethora of proper motion measurements from Hipparcos, it has become clear that
both a supernova ejection mechanism (where a star is shot out when its
companion explodes; Blaauw 1961) and a binary-binary collision mechanism
(where two binary systems collide and one star is scattered off through
many-body interactions \citep[e.g.][]{1992MNRAS.255..423C}) can produce so-called ``runaway" O and B stars,
whose velocities can exceed 100 \kms\ \citep {1986ApJS...61..419G, 2001A&A...365...49H}.

Amongst young and forming stars, the propensity for speeding is mostly
theoretical. Many simulations, of cluster \citep
[e.g.][]{1990AJ.....99..608L,2002MNRAS.336..705B, 2003MNRAS.339..577B,
2003MNRAS.343..413B}, small multiple \citep [e.g.][]{1995A&A...304L...9S,
1998MNRAS.301..759K}, and interacting binary \citep
[e.g.][]{1998MNRAS.298..231K} systems, show that large numbers of stars should be
ejected during the star formation process. However, most very young stars
are active, heavily embedded and/or associated with copious and variable
nebulosity, making radial velocity or proper motions measurements
challenging to impossible.

Nonetheless, PV Ceph is not the only young star to have been observed
speeding ---a handful of other young stars have also been justifiably accused (if not certainly convicted) of going
very fast.
Plambeck et al. (1995) show that the Becklin-Neugebauer object (BN) and the
radio source ``I" in the
Orion KL nebula are moving apart at $\sim 50$ \kms. From their data it is
not certain if one or
both objects
are moving.  However they suggest that BN is the more likely candidate, and
that it is probably a
runaway star
that formed in the Trapezium cluster.  The proper motion study of Loinard
(2002), using  two epochs of
VLA 3.6-cm continuum data, suggests that component A1 in the IRAS 16293-2422
protostellar system
could be a high-speed protostar with a velocity of $\sim 30$ \kms.   The
multi-epoch study of
Loinard, Rodr\'{\i}guez, \& Rodr\'{\i}guez (2003) proposed that a member of 
the T Tau triple system has recently been ejected and is now moving at $\sim 20$ \kms. 
But, the recent work of \citet{2003ApJ...596L..87F}
 shows that the radio-IR association made to reach
 this conclusion was likely in error, so we do not count this as a case of speeding.  In their study of proper motions of T Tauri variables in the
Taurus-Auriga complex,
Jones \& Herbig (1979) reported that RW Aur has a velocity of $\sim 16$
\kms\ on the plane of the
sky. We do not count this claim of speeding either, because more recent proper motion measurements of this star by the Hipparcos
and Tycho 2 missions show that the original measurements by Jones \& Herbig
(1979) are likely incorrect, and that the proper motion of RW Aur is within the typical
2 \kms \/ velocity dispersion of the cloud 
(L.~W.~Hartmann 2004, private communication). 
In cases more similar to PV Ceph's, where jet geometry is suggestive,
Bally, Devine, \& Alten (1996) suggest that motion at 5 to 10~km~s$^{-1}$
might explain the swept-back shape of the giant HH flow from B5-IRS1 in
Perseus.  And, Bally \& Reipurth (2001)
report the discovery of three jets in NGC 1333 cluster 
with a C-shaped symmetry,
similar to the B5-IRS1 outflow, with bending towards the cluster
core. They argue that the most likely scenario that explains the 
particular shape of these jets is one in which the jet sources have
been dynamical ejected from the cluster and now are
moving at about 10~km~s$^{-1}$ through the medium\footnote{The work of
Bally et al.~(1996) and Bally \& Reipurth (2001) 
give only rough estimates of the stellar velocities: they do
not employ a deceleration model like the one used
here for PV Ceph. If such a model were applied in the case of B5-IRS1, 
the velocity derived would be significantly larger.}.
Thus, at least seven young stars (PV
Ceph, BN/Source I, IRAS 16293-2422 (A1),
B5-IRS1, and three outflow sources in NGC 1333 (HH 334, HH 449, and HH 498))
have been accused at traveling at speeds in excess of 10 \kms\
through the interstellar medium.

Clearly, seven stars, all discovered to be speeders by serendipity, is not a
sample with which one can do statistics, so it is hard to say just how
globally common ``speeding" is amongst young stars.  Nonetheless, the fact
that seven speeders were caught without directed law-enforcement efforts
surely suggests that a focused campaign to measure the velocities of young
stars is now in order.  This campaign will not be easy with existing data.
To give an example, with a comparison 
of the original Palomar Survey (observed in 1952) and the Sloan Digital Sky
Survey (observed in 1995) we can place a $3-\sigma$ upper limit on PV Ceph's 
proper
motion of 33 \kms\ ---too large to constrain the source-motion scenario
proposed here.  Happily, the wealth of upcoming precision astrometry missions
(e.g. SIM, GAIA) should allow for a statistical sample of the motions of
young and forming stars.
 
\subsection{``Matching" Cores and Stars}

Ever since it was appreciated that stars form in the ``dense cores" of
molecular clouds \citep {1983ApJ...266..309M, 1986ApJ...307..337B}, it has
seemed puzzling that very few stars appear to be centered, alone in a blob
of dense gas.  Instead, it is
generally true that one finds many young stars where one finds copious
amounts of dense gas.  Sometimes, especially for
very young stars forming alone, it is easy to associate a star and its host
core, but even in those cases, the star is often
off-center with respect to gas features at $\sim 0.1$ pc scales.  In
clusters, it is nearly impossible to associate a single
star with a single core \citep [e.g.][]{1992ApJ...393L..25L}, and there is
also evidence that stars slowly diffuse out from the active part of the
cluster as they form \citep {1998ApJ...497..736H}.

The ``typical" velocity dispersion of molecular gas on 0.1 to 1 pc scales is
of order 1 \kms. Moving at 1 \kms, it would take a star 100,000 years to go
beyond the boundary of a core with radius 0.1 pc.  Given these numbers, one
can imagine a {\it slow} diffusion of stars out of their host cores, but not
many cases of way-off-center stars after just, the typical ``dynamical" time
of a (young!) HH flow (10,000 years).

If, instead, velocities of 1 \kms\ were still typical in the gas, but some
significant fraction of stars ---say those that had been ejected
from higher-order systems very early in their creation--- were moving at a
range of higher velocities, one could expect many failed
star-core associations.  One can even imagine estimating the typical stellar
velocity from star-core offsets ---but that's only possible if one can
identify a star's place of birth.  Surely, this was not an easy task for PV
Ceph!

\subsubsection{Estimating Star Formation Efficiency: What is the Right
Reservoir of ``Star-Forming" Gas?}
Star formation efficiency is typically calculated by dividing the stellar
output of some region by the gas mass in that region. How, though, should a
``region" be defined?  To take an extreme example, would PV Ceph and NGC 7023
have been thought of as from one ``region" ---likely not.

We suggest here that further thought needs to be given to (spatially) lost
generations of (still young) stars when calculating the star forming
efficiency of molecular clouds.  It may only be possible to do these
calculations on very large scales ---meaning that prior estimates of star
formation efficiency, especially of regional variations, may be misleading.
A similar suggestion has been made by Feigelson (1996) based on evidence for
a distributed population of weak-lined T-Tauri stars found far from their
birthplaces.  The evidence for the youth of these stars has been called into
question, but the suggestion that emigr\'ees need counting when calculating
star formation efficiency by census is surely correct.

\subsection{Nagging Questions}
\label{nagging}
If PV Ceph formed $\sim 10^{5}$ years ago in NGC 7023, how does it have such
an extensive molecular outflow, now?  Examining AGa,
Figure 5, one sees that ``first encounter"  of PV Ceph, moving at $\sim 20$
\kms, and the molecular cloud in which it now finds itself, would have been
about 35,000 years ago (0.7 pc to the East of its current position). As
shown by our analysis of the outflow in \S 2, which suggests that the
current features were created within the past 10,000 years, this is a
perfectly reasonable time constraint, but it does beg the question of what
PV Ceph was doing while it cruised through the vast expanse of nearly lower
density material between it and NGC 7023.  Did it start forming in the
cluster, then go dormant, and then get re-invigorated when it entered its
current host molecular cloud?

The Bondi-Hoyle analysis in \S \ref{groupies} implies that only a very small
blob ($\sim 0.001$ pc) of material could have traveled with PV Ceph from
NGC 7023 at 20 \kms.  Is it possible that PV Ceph took a disk with it, and
that the outflow/disk system just started ``working" (again?) as PV Ceph
entered its current molecular cloud?  One interesting point is that the
clockwise rotation of the outflow axis evident in Figure 1 implies that PV
Ceph's disk has changed its orientation over the past 10,000 years to be
parallel to its (nearly) E-W current direction of motion.\footnote{We credit
Dimitar Sasselov for causing us to wonder about the significance of
alignment of PV Ceph's disk and  its direction of motion.}  This could
easily be an accident, but investigating this ``coincidence" further is a
nagging question nonetheless.

\section{Summary}

While the evidence for PV Ceph moving at roughly 20 \kms\ is all
circumstantial, it comes from independent observations, and it is strong.
In \S \ref{plasmonmodel}, we show that a straightforward ``plasmon" model of
gas deceleration can readily explain the currently observed positions of the
optically-detected HH knots in the PV Ceph flow (see \S \ref{geometry}), but
only if PV Ceph itself is moving at roughly 22 \kms.  In \S \ref{trailsec}, we
show evidence for a trail of redshifted molecular gas, detected in both \co\
and \thco, at both low and high resolution, left behind in the wake of PV
Ceph as it moves along a mostly E-W path with a small component away from us.
Using HST observations of PV Ceph, we find the current jet direction to be
very close to N-S.  We compare this orientation with that of the wiggling
highest-velocity molecular jet observed at high resolution and find that the
offset between the optical and molecular jet directions is best explained if
PV Ceph is moving---at 21 \kms\ from E to W (\S \ref{wiggle}).  We emphasize
that the HST and IRAM molecular-line observations that give this velocity
are independent from the large-scale optical imaging that implies 22 \kms\
based on the HH knot geometry.

In Figure 7, we summarize the recent history of PV Ceph implied by our findings.  The Figure emphasizes that the knots in an HH flow from a young star can acquire a very noticeable velocity in the direction of the star's motion, if that star is moving rapidly.\footnote{Figure 7 is based on a movie prepared for a presentation of these results at the January 2004 meeting of the American Astronomical Society.  The movie showing the evolution of the knots in time, along with other information on PV Ceph, is available through: cfa$-$www.harvard.edu/$\sim$agoodman/Presentations/aas04PVCeph/.}

A velocity of $\sim 20$ \kms\ for a young star is unlikely to arise
randomly, and is much more likely caused by a dynamical ejection (\S
\ref{causes}).  In analyzing PV Ceph's environs, we find NGC 7023 ---10 pc
from PV Ceph in projection--- to be the nearest young cluster from which it
may have been ejected.  Astonishingly, we find that the back-track of PV
Ceph's motion, in all three spatial directions, seems to point quite
precisely to NGC 7023.  Furthermore, we find a dark rift in NGC 7023's
nebulosity that aligns eerily well with the exit path PV Ceph would have
taken from the cluster (Figure 6).

It would take PV Ceph hundreds of thousands of years to travel from NGC 7023
to its current location.  Most of the space in-between is of much lower
density than either NGC 7023 or the cloud now hosting PV Ceph.  What,
exactly, PV Ceph would have been doing during the trip (e.g. did it have a
disk? an outflow? was it dormant?) remains a difficult and fascinating, but
open question.

If it turns out that this seemingly bizarre story where PV Ceph was ejected
from NGC 7023 is right, we may need to reconsider the formation histories of
some other young stars, too.  To date, we know of seven examples, including PV
Ceph, of pre-main-sequence stars moving at $>>10$ \kms (\S \ref {speeders}).
All of these speeders have been caught serendipitously, and we suggest that
a systematic search for high-velocity young stars might yield dramatic
results.  If it turns out that speeding is a common offense ---as many modern
numerical simulations of the star formation process would
imply--- calculations of star formation efficiency in molecular clouds, and
perhaps even of the stellar initial mass function, will need to be
revisited.

\section{Acknowledgements}
This paper was initially intended to be a short ``Letter," based just on the
geometric arguments in \S \ref{geometry}.  The more people we talked with
about the work, though, the more interesting ---and more complex--- the
questions associated with it became.  The people responsible for the happy
expansion of this work include: John Bally, Ian Bonnell, Cathie Clarke, Bruce Elmegreen,
Lee Hartmann, Charles Lada, Richard Larson, Laurent Loinard, Lucas Macri, Douglas Richstone,
Phil Myers,  Ramesh Narayan, Eve Ostriker, Alex Raga, Bo Reipurth, Luis
Rodr\'{\i}guez, Edward Schwartz, Dimitar Sasselov, James Stone, and James Yorke.
We are deeply grateful to have had their thoughts, questions, and insights. 
We are also indebted to Deborah Padgett and Karl Staplefeldt for giving us
their team's HST images  of PV Ceph in advance of their publication
\citep{1999AAS...195.7902P}.  We thank the anonymous referee for pointing us to the highly relevant work on bent jets in clusters that we had somehow overlooked in our original submission.  AG is grateful to the Yale University
Astronomy Department for their kind hospitality while much of this work was
in progress. HGA would also like to acknowledge the partial support from NSF 
grant AST-9981546.




\clearpage


 \figcaption[f1.eps]{
The main (central) panel shows the integrated intensity of the $^{12}$CO(2-1) PV Ceph 
outflow (from AGa) superimposed on a wide-field H$\alpha$ + S[II] CCD
image (from Reipurth et al.~1997). Blue (red) contours denote the 
blueshifted (redshifted) lobe of the molecular outflow.
The position of the HH knots and that of PV Ceph are shown. 
The three 
colored, straight, lines (shown in all panels)
connect pairs of HH knots from the same ejection event, but opposing 
directions, from the star. The orange line connects the oldest ejection
event (the HH 315 C-F pair), the green line connects the HH 315 B-E event,
and the blue line connects the HH 315 A-D event.
The circle in each of the three colored lines
(shown in all panels) denotes the 
halfway point between the current positions of the given HH knot pair. 
The upper-left panel shows the redshifted $^{13}$CO(1-0) emission ($3.15 < v_{LSR} < 3.71$~\kms, from AGb)
superimposed on the same optical image as the main panel, 
but only covering a small ($2\arcmin \times 2\arcmin$) region surrounding PV Ceph. 
The middle-left panel shows the HST/WFPC2 image of the region around PV Ceph
(from Padgett, personal communication).
The two panels on the right show the high-resolution 
$^{12}$CO(2-1) redshifted outflow emission (from AGb). The upper panel shows
the ``less'' redshifted outflow integrated intensity ($3.16 < v_{LSR} < 4.15$~\kms) 
and the lower panel shows the ``more'' redshifted outflow integrated intensity 
($4.15 < v_{LSR} < 6.46$~\kms). The solid yellow line in these two panels shows the N-S position 
angle of the current PV Ceph outflow axis (see text).}

\figcaption[f2.eps]{
Plot showing the  loci of possible HH knot positions, in a reference frame tied to the moving source, using the plasmon model (see text). The (0,0) position corresponds to the current position of the
outflow source.
The different curves represent the results of the model
for different outflow source velocities (each curve is labeled with the velocity of the outflow
source). The rest of 
the parameter values for the three different curves shown are identical,
and are shown in Table 2.
The black dots represent the observed position of the
three different HH knot pairs of the HH 315 flow, in this coordinate system.
Notice that the model with a source velocity of 22~\kms\ (middle curve)
fits very well the observed knot positions in the PV Ceph flow.}

\figcaption[f3.eps]{
({\it Top.}) Plot of ``dynamical time'' over real elapsed time vs. the distance
of the HH knot from the emitting source. The dynamical time of an HH knot
is defined as the observed distance of the knot over the observed
velocity of the knot. ({\it Bottom.}) Plot of the velocity of the knot along
the jet direction vs. the distance of the HH knot from the emitting source. In both
panels the curve represents the results of the moving-source plasmon model (see text)
using the parameters in Table 2. The gray filled circles show where the
different HH knot pairs of the HH 315 flow would lie on the curves at the deprojected
distances of the observed knots. Each of the three HH knot pairs are labeled.}

\figcaption[f4.eps]{
Plot representing the amount of gas that might move along with a speeding protostar,
using the modified Bondi-Hoyle method (see text). Diagonal solid lines in the log-log
plot, show the relation between
$GM_{*}/R_{BH}$ and the Bondi-Hoyle accretion radius ($R_{BH}$) for
stars with different masses. The two dotted horizontal lines show where different
values of $\sigma_{eff}$ lie on the plot. The filled circle shows an estimate of
where PV Ceph would lie on the plot, with $\sigma_{eff} \sim $ (5~\kms)$^2$ and 
$M_{*} \sim 7$~M$_{\sun}$. The dashed vertical line indicates the estimated
value of $R_{BH}$ for PV Ceph ($R_{BH} \sim 0.001$~pc).}

\figcaption[f5.eps]{
({\it Left.}) Plot of the high-resolution $^{12}$CO(2-1) redshifted outflow lobe emission centroid position
(adapted from Figure 18 in AGb). The dark line shows the linear fit to the points. Distances
are given in terms of pixels and arcseconds from the map edge. The gray arrows show the
velocity vectors of the jet ($v_{jet}$) and of the star ($v_{*}$). ({\it Right.})  Same plot as
on the left panel, but corrected for the slope (linear fit) shown on the left panel. Distances are shown
in AU and arcseconds from the source. The sinusoidal fit to the points is also shown.}

\figcaption[f6.eps]{
Main (right) panel shows the 100\micron \/ IRAS image of the region near PV Ceph and the young stellar
cluster NGC 7023. Superimposed on the IRAS image, at the position of NGC 7023, is a DSS image of 
NGC 7023. This same DSS image is blown-up and shown on the upper-left panel. The magenta arrow in the 
main panel is an extension
towards the east of the presumed direction of motion of PV Ceph (from Figure 1), and corresponds
to the magenta line in the blown-up DSS image (upper-left panel). 
The white star symbols show the position of two IRAS sources
(possibly protostars) east of PV Ceph, IRAS 20495+6757 and IRAS 20514+6801. On the left panel, 
a dark lane in the
DSS image of NGC 7023 is identified as a possible gap produced by the ejection of PV Ceph from NGC 7023
(see text).} 

\figcaption[f7.eps]{Schematic illustration of PV Ceph's history.  For each pair of HH knots, small colored diamonds show the location, in time steps of 500 years, of the tip of a jet emitted from the point drawn as an ``explosion" of the same color. The relevant portions of Figure 1 are repeated in the background here, and the diamonds are color-coded to match the connecting-line for each pair of knots.  Larger diamonds emphasize the tips of those connecting lines, showing current HH knot positions. The plasmon model described in \S \ref{plasmonmodel} was used to create this figure, but the ejection angle for each pair of knots has been rotated by hand to match the observed positions of the knots.  Taken literally, this figure implies that the outermost pair of knots was emitted 9500 years ago, the middle pair 4700 years ago, and the innermost pair 2200 years ago, while PV Ceph traveled at 22 \kms.   A movie showing the evolution of the knots in time is available through: cfa$-$www.harvard.edu/$\sim$agoodman/Presentations/aas04PVCeph/.
} 


\clearpage

\begin{deluxetable}{lcccc}
\tablecolumns{5}
\tablecaption{Observed	Positions of Knots in the PV Ceph Flow
\label{pvcephknots}}
\tablehead{
\colhead {} & \colhead{Pair} & \colhead{} & \colhead{} & \colhead{}\\
\colhead{} & \colhead {Designation} & \colhead{$R_{knot}$\tablenotemark{a}} &
  \colhead{$R_{source}$\tablenotemark{b}} & \colhead{P.A.}\\
\colhead{} & \colhead{HH 315} & \colhead{[pc]} & \colhead{[pc]} & \colhead{[deg.~E of N]}
}
\startdata
Outermost Knots & C-F & 1.30 & 0.100 & 144\\
Middle Knots    & B-E & 0.90 & 0.056 & 156\\
Innermost Knots & A-D & 0.52 &  ---  & 159\\
\enddata

\tablenotetext{a}{$R_{knot}$ is one-half of the distance between the knot pair listed, measured
on the plane of the sky, assuming a distance to PV Ceph of 500~pc.}
\tablenotetext{b}{$R_{source}$ is the plane-of-the-sky distance between the bisector of a line
joining each knot pair listed and the current position of PV Ceph (which is the same, to within 
our measurement error, as the current position of the A-D bisector), assuming a distance to PV Ceph
of 500~pc.}
\tablenotetext{c}{P.A. gives the angle East of North defined by the line joining each pair of knots
(see Figure 1).}

\end{deluxetable}

\clearpage

\begin{deluxetable}{ll}
\tablecolumns{2}
\tablecaption{Parameters Used in Plasmon Model of the PV Ceph Flow
\label{plasmonmodeltab}}
\tablehead{}

\startdata
Star velocity, in direction of its motion\tablenotemark{a} & $v_{*} = 22$~\kms \\
Inclination of star's motion to sky plane\tablenotemark{b} & $\phi_{star-sky} = 20\arcdeg$\\
Initial velocity of ejected blobs\tablenotemark{a} & $v_{j} = 350$~\kms \\
Inclination of blob ejection (jet) axis to sky plane\tablenotemark{b} & $\phi_{flow-sky} = 45\arcdeg$\\
Initial angle between source motion and jet axis in $v_{j}-v_{*}$ plane\tablenotemark{c} & $\theta_{0} = 90\arcdeg$\\
Mass of each ejected blob\tablenotemark{b} & $M = 3 \times 10^{-4}$~M$_{\sun}$\\
Number density (of H$_{2}$) corresponding to $\rho_{amb}$\tablenotemark{c} & $n_{amb} = 1.50 \times 10^3$ cm$^{-3}$\\
Velocity dispersion in ambient medium\tablenotemark{c} (e.g., $^{12}$CO) & $\sigma = 3.2$~\kms \\
Geometric factor\tablenotemark{b} & $\xi = 14$\\
\enddata

\tablenotetext{a}{Fitted parameter value.}
\tablenotetext{b}{Reasonable assumption made for parameter value.} 
\tablenotetext{c}{Parameter value based on observations.}
\end{deluxetable}

\end{document}